\documentclass[superscriptaddress,amsmath,amssymb,aps,pra]{revtex4-2}

\usepackage{graphicx}
\usepackage{relsize} 
\usepackage{dcolumn}
\usepackage{bm}
\usepackage{amsmath} 
\usepackage[colorlinks]{hyperref}
\usepackage{braket}
\usepackage{cancel}
\usepackage{xcolor}

\begin{document}

\title{Anomalous Weak Pointer Shifts as Postselected Interference among Coarse-Grained Histories}

\author{Cody Charles Payne}
\affiliation{Department of Physics and Astrophysics, University of North Dakota, Grand Forks, North Dakota}

\author{Eliahu Cohen}
\affiliation{Faculty of Engineering and the Institute of Nanotechnology and Advanced Materials, Bar-Ilan University, Ramat Gan 5290002, Israel}
\affiliation{Institute for Quantum Studies, Chapman University, Orange, California 92866, USA}

\begin{abstract}
Anomalous weak values and superoscillations allow a pointer to exhibit an effective shift
outside the eigenvalue range of the measured observable. We reformulate this effect in
a phase-space path-integral language. For each eigenvalue branch $a_j$ of the measured
observable, {the interaction couples $a_j$ to the pointer position $\hat{x}_p$ and thereby translates the conjugate pointer momentum by $\gamma a_j$.} After
postselection, {the branch-conditioned amplitudes are coherently superposed with coefficients chosen to produce a superoscillatory approximation. When this conditional amplitude acts on an initial pointer momentum wavepacket for which the finite superoscillatory-window condition is satisfied, the resulting momentum wavefunction behaves as if it had been translated by $\gamma A_w$, rather than by any of the eigenvalue-conditioned shifts $\gamma a_j$. This yields an effective path-integral representation of anomalous weak pointer shifts and motivates their interpretation in terms of interference among finite coarse-grained histories.} We discuss how this picture bears on conservation-law
questions.

\end{abstract}

\maketitle

\section{Introduction}

An anomalous weak value \cite{spin-100,aharonov_ttm,aharonov2008quantum,dressel_rmp_review,aharonov_on_conservation_laws}, intimately related to the phenomenon of superoscillations \cite{aharonov_on_conservation_laws,Berry2019-xq}, refers to the value of a quantum observable within a pre- and post-selected system which is inferred from a shift in a pointer system weakly coupled to it that corresponds to a value not within the spectrum of the quantum system being measured. {Throughout this paper we use the momentum-readout convention: the measured observable is coupled to the pointer position $\hat{x}_p$, and the weak value is inferred from the resulting translation of the conjugate pointer momentum $\hat{p}_p$.}

{Values outside the spectral range have also been analyzed under the labels ``superweak'' \cite{berry_shukla_superweak} and ``strange'' weak values \cite{hosoya_shikano_strange}. Under suitable operational assumptions, anomalous weak values witness contextuality \cite{pusey_contextuality}, a connection that has been tested experimentally \cite{piacentini_contextuality}. It was also shown that anomalous weak values require that the pre- and postselection states have coherence in the eigenbasis of the measured observable \cite{wagner2023simple}. Finally, anomalous weak values have been demonstrated in a single-photon, single-detection protocol \cite{rebufello_single_photon,rebufello2025robust}.}

When the weak value being spoken of is the value of a conserved quantity such as momentum or energy as in \cite{aharonov_ttm,aharonov_on_conservation_laws}, this can raise interesting questions about the status of conservation laws within quantum mechanics and indeed this exact question is raised by Aharonov, Popescu, and Rohrlich (henceforth ``APR'') in \cite{aharonov_on_conservation_laws} and later given resolution in \cite{apr-conservation-laws-and-foundations}. However, this resolution leaves some room for further discussion of the ontology of conservation laws in that the resolution relies on a ``double non-conservation'' effect wherein the momentum (or energy) of a preparer system entangled with the pre- and post-selected quantum system receives an anomalous shift equal and opposite to that of the pointer. This entanglement itself arises as a result of macroscopic energy/momentum conservation. In light of this slight -- though not vicious -- bootstrapping effect, we feel it is valuable to consider conservation laws in this scenario on a slightly deeper level involving Feynman histories and path-integrals.  

The reason why an analysis  specifically in terms of path-integrals may be valuable may be seen by considering that within a path-integral, the amplitudes for \textit{all} possible (construed very broadly) histories must be considered and integrated over, including {nonstationary histories that need not satisfy the classical
equations of motion.} Therefore an initial motivating question arises: could scenarios of the pointer receiving anomalous shifts in these conserved quantities correspond to a family of off-shell Feynman histories which typically lie far outside the path of stationary action and thus destructively interfere within the path-integral?  
{The primary thrust of this paper is to suggest an affirmative answer
to this question. We derive a model in which pre- and postselection
modify the effective phase of the branch-summed conditional amplitude
so that a coarse-grained tube centered on the anomalous pointer
trajectory contains a stationary region and can therefore contribute
coherently. Related interpretations have been proposed, e.g., by
Aharonov and Rohrlich \cite{aharonov2008quantum} and by Addazi and Gan
\cite{weak_values_in_QFT}. Here we place this idea within a more general framework.}

In a weak-measurement setup with postselection, the pointer momentum wavefunction can be written
as a coherent superposition of ordinary eigenvalue-conditioned momentum translations. In the
superoscillatory regime, this superposition acts on the initial momentum wavepacket as an
effective translation in which the eigenvalue is replaced by the weak value. The corresponding
phase-space path-integral gives an effective conditional representation of that momentum shift
and motivates an interpretation in terms of finite coarse-grained momentum histories.

{Related analyses emphasize the same phase and interference structure: Hofmann connected the phases of weak conditional probabilities with the action associated with unitary transformations \cite{hofmann_complex_phases}, while Dressel formulated weak values explicitly as interference phenomena \cite{dressel_interference}.}

This work is closest in spirit to Georgiev and Cohen \cite{cohen_probing_virtual_histories_with_weak_values}, who analyze finite coarse-grained
virtual Feynman histories using sequential weak values, and to Matzkin \cite{matzkin,Matzkin_2015}, who derives weak
values from path-integrals more generally. The present paper focuses on a more specific
mechanism: the superoscillatory weak-pointer shift. We show how the postselected pointer
amplitude can be written as a superposition of eigenvalue-conditioned path-integral
kernels and how, within the superoscillatory window, this superposition behaves as an
effective anomalous momentum-history contribution. The resulting picture is then used
to motivate, rather than fully prove, a connection to conservation-law debates. In particular, {we argue that this representation helps distinguish anomalous
conditional behavior of a subsystem from violation of a conserved
total charge, yet  it does not, by itself, constitute a complete treatment
of the APR preparation problem.}

This paper is organized as follows. First we discuss weak values generally as they are typically derived in the literature, without reference to path-integrals. We then derive the effective path-integral kernel (propagator) given the correct (superoscillatory) pre- and post-selection procedure and how it results in equations of motion that favor {momentum histories centered on the anomalous weak value shift}. We then give explicit form to the ontology we are proposing in terms of coarse-grained history families using the history projection operator formalism. Finally, we briefly discuss conservation law puzzles in light of this ontology, without attempting a complete treatment, which we leave to future work. In particular, this treatment in terms of path-integrals naturally suggests a rigorous approach towards conservation within the full system-pointer-preparer setting in terms of Ward identities.

\section{Weak Values}

The weak value $A_w$ of a quantum observable $\hat{A}$ is a value that may be assigned to that observable between two measurements, namely the measurement that pre-selects the quantum system in the state $|a_i\rangle$ and the measurement which post-selects it in the state $|a_f\rangle$. {Its value is given by \cite{spin-100}}:

\begin{align}
    A_w=\frac{\langle a_f|\hat{A}|a_i\rangle}{\langle a_f|a_i\rangle}.
\end{align}

It is of course possible to express the pre- and post-selected states as superpositions of eigenstates of $\hat{A}$, namely $\left\{|a_j\rangle\right\}$ where $\hat{A}|a_j\rangle=a_j|a_j\rangle$.

{It is useful to consider the weak value in the momentum-readout version of a von Neumann interaction. The measured-system observable $\hat{A}_s$ is coupled to the pointer position $\hat{x}_p$, so that the conjugate pointer momentum $\hat{p}_p$ is translated:}

\begin{align}
    \hat{H}_{\mathrm{int}}=-g(t)\hat{A}_s\hat{x}_p.
\end{align}

{Here $[\hat{x}_p,\hat{p}_p]=i\hbar$. The subscripts $s$ and $p$ label the measured system and the pointer, respectively. When no ambiguity can arise, we suppress the pointer subscript, writing $\hat{x}_p\to\hat{x}$ and $\hat{p}_p\to\hat{p}$ for operators, and $x_p\to x$ and $p_p\to p$ for the corresponding phase-space variables.} The function $g(t)$ parametrizes the time-dependent coupling between system and pointer and it is the case that:

\begin{align}
    \int_{t_i}^{t_f}\mathrm{d}t\,g(t)\overset{\text{def}}{=}\gamma.
\end{align}

{For a branch with $\hat{A}_s|a_j\rangle=a_j|a_j\rangle$, the impulsive pointer unitary and its action on the pointer momentum are}
\[
    {\hat{U}_j=\exp\!\left(\frac{i}{\hbar}\gamma a_j\hat{x}_p\right),\qquad
    \hat{U}_j^{\dagger}\hat{p}_p\hat{U}_j=\hat{p}_p+\gamma a_j.}
\]
{Thus $\hat{x}_p$ is the pointer variable appearing in the coupling, whereas $\hat{p}_p$ is the readout variable that acquires the shift.} The parameter $\gamma$ (which we will typically assume to be weak relative to the pointer momentum resolution) defines the overall strength of the interaction. {We therefore specify the initial pointer state primarily in the momentum representation, $\widetilde{\phi}_i(p)=\langle p|\phi_i\rangle$. For a branch corresponding to $|a_j\rangle$, the unnormalized postselected momentum wavefunction is}

\begin{align}
    {\widetilde{\Phi}^{(j)}_f(p)}
    &{=\left\langle p\right|\left(\langle a_f|\hat U_j|a_j,\phi_i\rangle\right)}\nonumber\\
    &{=\langle a_f|a_j\rangle\,\widetilde{\phi}_i\!\left(p-\gamma a_j\right).}
\end{align}

{Thus branch $j$ translates the pointer momentum distribution by $+\gamma a_j$. In the Fourier-conjugate position representation the same translation operator is multiplication by $\exp(i\gamma a_jx/\hbar)$. We use that auxiliary representation below only to display the superoscillatory phase relation; the measured readout throughout remains the pointer momentum.}

{
One of the important insights of Aharonov et al. \cite{spin-100,aharonov_ttm, aharonov_on_conservation_laws,apr-conservation-laws-and-foundations,aharonov2008quantum} is that it is possible to choose coefficients $C_j$ for a preselected state

\begin{align}
    |a_i\rangle=\sum_{j=0}^{N}C_j|a_j\rangle
\end{align}

such that the exact unnormalized postselected pointer state in the momentum representation is

\begin{align}
    \widetilde{\Phi}_f(p)
    =\sum_{j=0}^{N}C_j\,\langle a_f|a_j\rangle\,
    \widetilde{\phi}_i\!\left(p-\gamma a_j\right).
\end{align}

Here $\Lambda$ is an overall normalization/postselection amplitude. A concrete construction is

\begin{equation}
\begin{aligned}
    c_j &\equiv
    2^{-N}\binom{N}{j}(1+\beta)^{N-j}(1-\beta)^j,\\
    C_j &\equiv
    \Lambda\,\frac{c_j}{\langle a_f|a_j\rangle}.
\end{aligned}
\end{equation}

Here $a_j=\left(1-2j/N\right)a_0$, $N$ is even, $\beta>1$, and the construction yields $A_w=\beta a_0$. We assume $\langle a_f|a_j\rangle\neq0$ for every branch retained in the sum. Choosing $\Lambda$ so that $|a_i\rangle$ is normalized gives, up to an irrelevant overall phase,
\[
    |\Lambda|^{-2}
    =
    \sum_{j=0}^{N}
    \frac{|c_j|^2}{|\langle a_f|a_j\rangle|^2}.
\]

The superoscillatory mechanism is most transparent in the Fourier-conjugate position variable $x$, because $x$ generates translations of the momentum readout. With these coefficients, the translation multiplier is exactly

\begin{align}\label{so_phase}
    \sum_{j=0}^{N}C_j\,\langle a_f|a_j\rangle\,
    \exp\!\left(\frac{i}{\hbar}\gamma a_j x\right)
    &=
    \Lambda\left(
    \cos\frac{\gamma a_0 x}{N\hbar}
    +i\beta\sin\frac{\gamma a_0 x}{N\hbar}
    \right)^N
    \nonumber\\
    &\approx
    \Lambda\exp\!\left(\frac{i}{\hbar}\gamma A_w x\right).
\end{align}

The final approximation is local in the conjugate variable $x$. Writing $y\equiv\gamma a_0x/\hbar$, sufficient local conditions are $|y|/N\ll1$ and $(\beta^2-1)y^2/(2N)\ll1$ over the effective support of the Fourier transform $\phi_i(x)=\langle x|\phi_i\rangle$ of the initial momentum wavepacket. Fourier transforming Eq.~\eqref{so_phase} gives the operational momentum-readout statement

\begin{align}\label{so_eqn}
    \widetilde{\Phi}_f(p)
    \approx
    \Lambda\,\widetilde{\phi}_i\!\left(p-\gamma A_w\right).
\end{align}

Equation~\eqref{so_eqn} says that the postselected momentum distribution behaves as though it were translated by $+\gamma A_w$. The appearance of $x$ in Eq.~\eqref{so_phase} specifies the Fourier-domain window in which this translation approximation is valid; it does not change the choice of momentum as the pointer readout. Contributions from the tails of $\phi_i(x)$ outside that window are neglected, and the anomalous effective shift is accompanied by a normalization cost reflected in $\Lambda$ and in the postselection probability.

In the path-integral treatment below, the same approximation is applied to a coarse-grained class of paths for which the interaction functional $X[x]\equiv\int_{t_i}^{t_f}g(t)x(t)\,\mathrm{d}t$ remains inside the superoscillatory window. In the impulsive limit, $x(t)$ is effectively constant during the interaction and $X[x]\simeq\gamma x$. With this qualification, the weak action is an effective conditional phase functional for the branch-summed, postselected momentum amplitude.
}

One of the more notable aspects of $A_w$ is that it does \textit{not} need to be within the eigenspectrum of $\hat{A}$, and in fact can be arbitrarily larger (or smaller) than any of the values $a_j$.  {In general, $A_w$ may be complex, but in this manuscript we restrict
attention to real weak values. Here ``anomalous'' means that the real
value $A_w$ lies outside the eigenvalue range of $\hat A$. In the convention used here, the measured system is coupled to the pointer position, while the readout is the induced shift of the conjugate pointer momentum. It is this anomalous momentum shift that naturally raises questions about conservation laws.}

{As a direct wavepacket-level check of the momentum-readout approximation, let the initial pointer momentum wavefunction be the normalized Gaussian
\begin{equation}
\widetilde{\phi}_i(p)
=
(2\pi\sigma_p^2)^{-1/4}
\exp\!\left[-\frac{(p-p_i)^2}{4\sigma_p^2}\right].
\end{equation}
Using $C_j\langle a_f|a_j\rangle=\Lambda c_j$, the exact unnormalized postselected momentum wavefunction after the impulsive interaction is
\begin{equation}
\widetilde{\Phi}_f(p)
=
\Lambda\sum_{j=0}^{N}c_j\,
\widetilde{\phi}_i\!\left(p-\gamma a_j\right),
\end{equation}
whereas the effective weak value description predicts
\begin{equation}
\widetilde{\Phi}^{(w)}_f(p)
=
\Lambda\,\widetilde{\phi}_i\!\left(p-\gamma A_w\right).
\end{equation}
A direct measure of the validity of the effective momentum translation is
\begin{equation}
\epsilon_{\mathrm{so}}
=
\frac{
\left\|
\sum_{j=0}^{N}c_j\,
\widetilde{\phi}_i(\,\cdot-\gamma a_j)
-
\widetilde{\phi}_i(\,\cdot-\gamma A_w)
\right\|_2
}{
\|\widetilde{\phi}_i\|_2
}.
\end{equation}
Equivalently,
\begin{equation}
\epsilon_{\mathrm{so}}^2
=
\frac{
\displaystyle
\int_{-\infty}^{\infty}\mathrm{d}p\,
\left|
\sum_{j=0}^{N}c_j\,
\widetilde{\phi}_i(p-\gamma a_j)
-
\widetilde{\phi}_i(p-\gamma A_w)
\right|^2
}{
\displaystyle
\int_{-\infty}^{\infty}\mathrm{d}p\,
|\widetilde{\phi}_i(p)|^2
}.
\end{equation}
For the Gaussian above, the Fourier-conjugate position width is $\sigma_x=\hbar/(2\sigma_p)$. Hence the dimensionless parameter controlling how much of the conjugate position profile lies in the superoscillatory window is
\begin{equation}
\eta
\equiv
\frac{\gamma a_0\sigma_x}{\hbar}
=
\frac{\gamma a_0}{2\sigma_p}.
\end{equation}
The dependence of $\epsilon_{\mathrm{so}}$ on $N$, $\beta$, and $\eta$ provides a quantitative test of the superoscillatory approximation for the actual momentum wavepacket. The postselection cost should be monitored separately through
\begin{equation}
P_{\mathrm{post}}
=
\|\widetilde{\Phi}_f\|_2^2
=
|\Lambda|^2
\int_{-\infty}^{\infty}\mathrm{d}p\,
\left|
\sum_{j=0}^{N}c_j\,
\widetilde{\phi}_i(p-\gamma a_j)
\right|^2.
\end{equation}
When the superoscillatory approximation is accurate and $A_w$ is real,
\begin{equation}
P_{\mathrm{post}}
\approx
|\Lambda|^2,
\end{equation}
because a real momentum translation preserves the norm. The effective state has mean momentum
\begin{equation}
\frac{
\displaystyle
\int_{-\infty}^{\infty}\mathrm{d}p\,
p\,\left|\widetilde{\Phi}^{(w)}_f(p)\right|^2
}{
\displaystyle
\int_{-\infty}^{\infty}\mathrm{d}p\,
\left|\widetilde{\Phi}^{(w)}_f(p)\right|^2
}
=
p_i+\gamma A_w,
\end{equation}
which is the anomalous momentum readout studied in the remainder of the paper.}

\section{Weak Values in the Path-Integral Framework}

Consider the classical Lagrangian associated with the pointer in phase space for a given branch associated with initial condition $|a_j\rangle$ and for fixed momentum endpoints $p_i,p_f$. {Here $x(t)$ denotes the pointer position and $p(t)$ its canonically conjugate momentum:}

\begin{align}
    \mathcal{L}^{(j)}[x,p]=-\dot{p}x-H_0-H_{\mathrm{int}}=-\dot{p}x-\frac{p^2}{2m}+g(t)a_j x
\end{align}

With an associated action functional:

\begin{align}
    S^{(j)}[x,p]=\int_{t_i}^{t_f}\mathrm{d}t\,\mathcal{L}^{(j)}[x,p]=\int_{t_i}^{t_f}\mathrm{d}t\,\left[-\dot{p}x-\frac{p^2}{2m}+g(t)a_j x\right]\nonumber\\\nonumber\\
    \rightarrow \delta S^{(j)}=\int_{t_i}^{t_f}\mathrm{d}t\,\left[\left\{\partial_x\mathcal{L}^{(j)}-\mathrm{d}_t\left(\partial_{\dot{x}}\mathcal{L}^{(j)}\right)\right\}\delta x+\left\{\partial_p\mathcal{L}^{(j)}-\mathrm{d}_t\left(\partial_{\dot{p}}\mathcal{L}^{(j)}\right)\right\}\delta p\right]\nonumber\\\nonumber\\
    =\int_{t_i}^{t_f}\mathrm{d}t\,\left[\left(-\dot{p}+g(t)a_j\right)\delta x+\left(\dot{x}-\frac{p}{m}\right)\delta p\right]
\end{align}

{Here $x(t)$ and $p(t)$ are independent phase-space variables. With fixed momentum endpoints, the boundary term $-x\,\delta p|_{t_i}^{t_f}$ vanishes. Thus varying $x$ gives the equation for its conjugate momentum $p$, exactly as in the standard first-order Hamiltonian variational principle; no higher-derivative Euler--Lagrange equation is required.}

Applying Hamilton's principle $(\delta S=0)$ and noting that $x$ and $p$ vary independently  gives equations of motion:

\begin{align}
    \dot{p}=g(t) a_j\rightarrow p(t)=p_i+G(t)a_j,\;\;\dot{x}=\frac{p}{m},
\end{align}

where:

\begin{align}
    G(t)=\int_{t_i}^t\mathrm{d}t'\,g(t').
\end{align}

Therefore the path in momentum space which extremizes the action given a branch $j$ is that path such that $p(t)=p_i+G(t)a_j$, as one would expect. 

{For a fixed eigenvalue branch $j$, define the momentum-endpoint
phase-space kernel
\begin{equation}
K_j(p_f,t_f;p_i,t_i)
=
\int_{p(t_i)=p_i}^{p(t_f)=p_f}
\mathcal Dp\,\mathcal Dx\,
\exp\!\left[
\frac{i}{\hbar}S^{(j)}[x,p]
\right].
\end{equation}
Because $x$ enters the branch action linearly, its functional
integration gives
\begin{align}
K_j(p_f,t_f;p_i,t_i)
&\propto
\int\mathcal Dp\,
\delta\!\left[\dot p-g(t)a_j\right]
\exp\!\left[
-\frac{i}{\hbar}
\int_{t_i}^{t_f}\frac{p(t)^2}{2m}\,dt
\right].
\end{align}
Thus the branch kernel is supported on
\begin{equation}
p_j(t)=p_i+G(t)a_j,
\end{equation}
and, for exact momentum endpoints, contains a factor proportional to
\begin{equation}
\delta\!\left(p_f-p_i-\gamma a_j\right).
\end{equation}}

{The superoscillatory approximation is therefore not a pointwise
identity between the bare momentum kernels.  It is an approximation
to the action of the postselected kernel on the initial pointer
wavepacket.  In the impulsive regime,
\begin{equation}
\Phi_f(x)
=
\phi_i(x)
\sum_j C_j\langle a_f|a_j\rangle
\exp\!\left(\frac{i}{\hbar}\gamma a_jx\right)
\approx
\Lambda\,
\exp\!\left(\frac{i}{\hbar}\gamma A_wx\right)
\phi_i(x)
\end{equation}
over the effective support of $\phi_i(x)$ inside the
superoscillatory window.}

{On this restricted class of pointer states, the branch-summed
conditional amplitude may be represented by the effective action
\begin{equation}
S^{(w)}[x,p]
=
\int_{t_i}^{t_f}\mathrm{d}t\,
\left[
-x\dot p-\frac{p^2}{2m}+g(t)A_wx
\right].
\end{equation}
The stationary condition obtained by varying the effective action
with respect to $x(t)$ is
\begin{equation}
\frac{\delta S^{(w)}}{\delta x(t)}
=
-\dot p+g(t)A_w
=
0.
\end{equation}
Consequently, the effective conditional trajectory is centered on
\begin{equation}
p_w(t)=p_i+G(t)A_w,
\qquad
p_w(t_f)=p_i+\gamma A_w.
\end{equation}}

{If the effective exponential were formally extended over the entire
$x$-domain, integration over $x$ would give
$\delta[\dot p-g(t)A_w]$.  In the actual problem, however, the
superoscillatory replacement is valid only within a finite window of
the functional $X[x]$.  The corresponding functional integral is
therefore peaked around $\dot p=g(t)A_w$, rather than being an exact
delta functional.  Accordingly, $p_w(t)$ should be interpreted as the
stationary center of an enhanced coarse-grained tube of histories,
not as the exact support of a globally valid effective kernel.  For
fixed momentum endpoints, this tube contributes appreciably only
when the final coarse-graining window contains
$p_i+\gamma A_w$.
}

The weak value history is not one of the eigenvalue-conditioned classical pointer histories. It is an effective conditional history of the postselected amplitude, valid in the
superoscillatory window in which the coherent sum of eigenvalue branches behaves as a
single anomalous exponential.

{Here ``constructive'' and ``destructive'' interference are meant in the stationary-phase sense, not as exact pairwise cancellation. After time slicing, a coarse-grained path-integral has the form $\int_{\Omega}\mathrm{d}^n z\,a(z)\exp[iS(z)/\hbar]$. On a region where $S$ has no stationary point and the weight $a$ is sufficiently regular, the factor $1/\hbar$ makes the phase rapidly oscillatory whenever the dimensionless action variation satisfies $|\delta S|/\hbar\gg 1$. The Riemann--Lebesgue principle, together with standard non-stationary-phase estimates obtained by integration by parts, then motivates suppression of that region. Neighborhoods in which $\nabla S=0$ are not suppressed by this mechanism and give the leading stationary-phase contribution. Strictly, the Riemann--Lebesgue lemma applies to ordinary finite-dimensional integrals, so its use here is understood through the time-sliced regularization of the path-integral. Our claim is therefore comparative: within its validity window, the superoscillatory branch sum changes the effective conditional phase $S^{(w)}$ so that the coarse-grained tube centered on $p_w(t)=p_i+G(t)A_w$ contains a stationary region. This does not imply exact cancellation outside the tube.}

\section{Interpretation in Terms of Coarse-Grained Histories }

{Let
\begin{equation}
\hat P_{\Delta_m}
=
\int_{\Delta_m}\mathrm{d}\mathcal P\,
|\mathcal P\rangle\langle\mathcal P|
\end{equation}
be a Schr\"odinger-picture projector onto a finite momentum interval,
where
\begin{equation}
\Delta_m
=
\left[
p_w(t_m)-\Delta,\,
p_w(t_m)+\Delta
\right],
\qquad
p_w(t_m)=p_i+G(t_m)A_w.
\end{equation}
The class operator associated with the corresponding coarse-grained
tube $\alpha$ is
\begin{align}
\hat C_\alpha
={}&
\hat U(t_f,t_M)
\left(\hat I_s\otimes\hat P_{\Delta_M}\right)
\hat U(t_M,t_{M-1})
\cdots \nonumber\\
&\times
\left(\hat I_s\otimes\hat P_{\Delta_1}\right)
\hat U(t_1,t_i),
\end{align}
where the ordering has been written explicitly.}

{After postselection of the measured system, the class operator defines
an unnormalized pointer state
\begin{equation}
|\Phi_\alpha\rangle_p
=
\langle a_f|\hat C_\alpha|a_i,\phi_i\rangle.
\end{equation}
Under the same branch-preserving assumptions used above, its momentum
wavefunction has the schematic path-integral representation
\begin{align}
\widetilde{\Phi}_\alpha(p_f)
\propto
\sum_j C_j\langle a_f|a_j\rangle
\int_{-\infty}^{\infty}\!\mathrm{d}p_0\,
\widetilde{\phi}_i(p_0)
\int_{\substack{p(t_i)=p_0\\ p(t_f)=p_f}}
\mathcal Dp\,\mathcal Dx\,
\chi_\alpha[p]\,
\exp\!\left[
\frac{i}{\hbar}S^{(j)}[x,p]
\right].
\end{align}
Here $p_0$ is a dummy integration variable over the initial momentum;
$p_i$ continues to denote the central momentum of the initial Gaussian.
The factor
\begin{equation*}
\mathbf 1_{\Delta_m}(q)
=
\begin{cases}
1, & q\in\Delta_m,\\
0, & q\notin\Delta_m,
\end{cases}
\end{equation*}
is the indicator (or characteristic) function of the interval
$\Delta_m$.  Consequently,
\begin{equation}
\chi_\alpha[p]
=
\prod_{m=1}^{M}
\mathbf 1_{\Delta_m}\!\left(p(t_m)\right)
\end{equation}
equals one precisely when the path satisfies
$p(t_m)\in\Delta_m$ at every selected time $t_m$, and equals zero
otherwise.  It therefore removes from the path-integral every path
that leaves at least one of the prescribed momentum windows.}

{For the tube $\alpha_w$ centered on the weak value trajectory, the
superoscillatory approximation gives
\begin{align}
\widetilde{\Phi}_{\alpha_w}(p_f)
\approx
\Lambda
\int_{-\infty}^{\infty}\!\mathrm{d}p_0\,
\widetilde{\phi}_i(p_0)
\int_{\substack{p(t_i)=p_0\\ p(t_f)=p_f}}
\mathcal Dp\,\mathcal Dx\,
\chi_{\alpha_w}[p]\,
\exp\!\left[
\frac{i}{\hbar}S^{(w)}[x,p]
\right],
\end{align}
provided that the relevant values of $X[x]$ remain inside the
superoscillatory window.  This is the momentum-space Green-function
construction: the kernel is propagated from each possible initial
momentum $p_0$ and weighted by $\widetilde{\phi}_i(p_0)$.  No separate
$\mathrm{d}x_i$ integral is required, because the initial state is
specified in the momentum representation and the conjugate variable
$x(t)$ is already integrated over as part of $\mathcal Dx$.}

{The stationary condition for $S^{(w)}$ is centered on
$p_w(t)=p_i+G(t)A_w$.  A sufficiently narrow tube centered instead
on an eigenvalue-conditioned trajectory
$p_k(t)=p_i+G(t)a_k$, and disjoint from the weak value tube, contains
no stationary trajectory of the effective conditional action and is
therefore expected to be relatively suppressed by nonstationary-phase
cancellation.  This is a comparative statement about class
amplitudes.  We do not claim that the relevant amplitudes are
numerically equal to $0$, $1$, or $\Lambda$; their magnitudes depend
on the initial pointer wavepacket, the postselection probability, the
window widths, and boundary contributions.}

{We use these class operators only to decompose the postselected
transition amplitude into finite coarse-grained alternatives.  No
consistent-histories probability is assigned unless the corresponding
decoherence functional is approximately diagonal or an explicit
measurement protocol is supplied.}

{This becomes relevant in APR-type puzzles because an individual off-shell history need not satisfy the classical equations of motion or the classical Noether conservation equation pointwise. Conservation at the quantum level is instead encoded in a symmetry of the full action and measure. For an infinitesimal change of variables, and assuming that the boundary states are transformed consistently and that there is no anomalous Jacobian, the corresponding pre- and postselected Ward identity may be written schematically as}

\begin{align}
    {0=\langle\delta\mathcal{O}\rangle_{fi}+\frac{i}{\hbar}\langle\mathcal{O}\,\delta S\rangle_{fi},}
\end{align}

{where $\langle\cdot\rangle_{fi}$ denotes the normalized conditional path-integral average and possible boundary terms have been left implicit. For a continuous symmetry generated by the total charge $Q_{\mathrm{tot}}$, the equivalent operator statement is}

{
\begin{align}
Q_{\mathrm{tot},w}(t)
&=
\frac{
\langle f|
U(t_f,t)Q_{\mathrm{tot}}(t)U(t,t_i)
|i\rangle
}{
\langle f|U(t_f,t_i)|i\rangle
},
\nonumber\\
\frac{\mathrm{d}}{\mathrm{d}t}Q_{\mathrm{tot},w}(t)
&=
\left(
\frac{\partial Q_{\mathrm{tot}}}{\partial t}
+
\frac{i}{\hbar}
[H_{\mathrm{tot}},Q_{\mathrm{tot}}]
\right)_w
=
0
\end{align}
whenever $Q_{\mathrm{tot}}$ satisfies the corresponding quantum
conservation equation.  Thus, the path-integral picture permits
nonstationary contributions without relaxing conservation of the
total charge in the full closed-system amplitude.}

{For energy conservation, the relevant closed system must be
autonomous.  The prescribed time-dependent function $g(t)$ in the
pointer Hamiltonian treats the source and timing of the interaction
externally, so the reduced pointer model cannot by itself establish
conservation of total energy.  A complete treatment must include the
degrees of freedom that generate and switch the interaction, together
with the measured system, pointer, preparer, and relevant reference
or clock degrees of freedom.  An anomalous weak value of one subsystem
is not by itself a violation of the total Ward identity.}

\section{Discussion}

{We have provided a path-integral framework for interpreting anomalous
weak pointer shifts in terms of non-classical, coarse-grained
histories. Such histories are not created by the pre- and
post-selection: their contributions are already present in the sum
over histories, although a tube centered on the anomalous trajectory
is generally nonstationary for each ordinary eigenvalue branch and
its contributions therefore tend to cancel. The superoscillatory
pre- and post-selection coherently recombines the
eigenvalue-conditioned branches so that, within the relevant window,
the postselected amplitude is described by an effective phase whose
stationary region is centered on the anomalous weak-value trajectory.
In this sense, the interference landscape is reshaped: a
family of histories that would ordinarily be strongly suppressed can
acquire an appreciable conditional amplitude, while coarse-grained
tubes disjoint from the new stationary region are relatively
suppressed. This provides an intuitive picture of how the anomalous
shift emerges and helps clarify why unusual conditional behavior of
the pointer need not amount to a violation of conservation by the
complete closed system.}

The present analysis should not be read as a complete resolution of the APR conservation law problem. What it provides is a path-integral representation of anomalous weak
pointer shifts as postselected interference among coarse-grained histories. In such
a representation, individual off-shell histories are not required to obey the classical
conservation law. The relevant conservation statement applies to the full closed-system
amplitude, or equivalently to the weak value of the total conserved generator. A complete treatment of APR-type conservation-law puzzles would therefore require including the
preparer and reference degrees of freedom and deriving the corresponding Ward identity
or total weak value conservation relation.

This framework may also be seen as providing some evidence for an \textit{ontological} as opposed to merely \textit{epistemic} view of the path-integral approach to quantum mechanics. If weak values are viewed as revealing non-classical Feynman histories -- and indeed this is in line with some previous literature on the subject, most notably \cite{cohen_probing_virtual_histories_with_weak_values} -- then the weak values in turn provide some evidence that these non-classical histories are in some sense ``real'', where they are typically viewed only as an artifact of a clever mathematical ``trick''. 

\section*{Acknowledgements}
This work was supported by the European Union’s Horizon Europe research and innovation programme under grant agreement No. 101178170, by the Israel Science Foundation under grant agreement No. 2208/24, by the Israel Innovation Authority and by the Israeli Council for Higher Education through ``QERNEL'' and the ``Interdisciplinary Center for the Theory of Quantum Computing''.


\bibliography{citations} 
\end{document}